\documentclass[a4paper,fleqn,usenatbib]{mnras}

\usepackage[T1]{fontenc}
\usepackage{ae,aecompl}

\usepackage{graphicx}	
\usepackage{amsmath}	
\usepackage{amssymb}	

\title[Structure of the gravitoturbulent quasar  discs]{Structure of radiation dominated gravitoturbulent quasar discs}

\author[M. Shadmehri, F. Khajenabi \& S. Dib]{
Mohsen Shadmehri$^1$,\thanks{E-mail: m.shadmehri@gu.ac.ir}
Fazeleh Khajenabi$^1$,
Sami Dib$^{2,3}$ 
\\
$^1$ Department of Physics, Faculty of Sciences, Golestan University, Gorgan 49138-15739, Iran\\
$^{2}$Niels Bohr Institute \& Centre for Star and Planet Formation, University of Copenhagen, {\O}ster Voldgade 5-7, DK-1350, Copenhagen, Denmark.\\ 
$^{3}$Unidad de Astronom\'{i}a, Departamento de Fisica, Universidad de Atacama, Copayapu 485, Copiapo, Chile.\\
}

\date{Accepted XXX. Received YYY; in original form ZZZ}

\pubyear{2015}

\begin{document}
\label{firstpage}
\pagerange{\pageref{firstpage}--\pageref{lastpage}}
\maketitle

\begin{abstract}
Self-gravitating accretion discs in a  gravitoturbulent state, including radiation and gas pressures,  are studied using a set of new analytical solutions. While the Toomre parameter of the disc remains close to its critical value for the onset  of gravitational instability,  the dimensionless stress parameter is uniquely determined from the thermal energy reservoir of the disc and its cooling rate. Our solutions are applicable to the accretion discs with dynamically important radiation pressure like in the quasars discs. We show that physical quantities  of a gravitoturbulent disc in the presence of radiation are  significantly modified compared to solutions with only gas pressure. We show that the dimensionless stress parameter is an increasing function of the radial distance so that its steepness  strongly depends on the accretion rate.  In a  disc without radiation its slope is 4.5, however, we show that in the presence of radiation, it varies between 2 and 4.5 depending on the accretion rate and the central mass. As for the surface density, we find a shallower profile with an exponent -2 in a disc with sub-Eddington accretion rate compared to a similar disc, but without radiation,  where its surface density slope is -3 independent of the accretion rate. We then investigate gravitational stability of the disc when the stress parameter reaches to its critical value. In order to self-consistently determine the fragmentation boundary, however, it is shown that the critical value of the stress parameter  is a power-law function of the ratio of gas pressure and the total pressure and its exponent is around 1.7.  We also estimate the maximum mass of the central black hole using our analytical solutions.
\end{abstract}




\section{Introduction}

Accretion discs around supermassive black holes (SMBHs) at the center of almost every galaxy are believed to be the main source of their huge luminosities.   This intense radiation allows us to detect them  as quasars and active galactic nuclei (AGNs). Enormous efforts have gone into understanding the gravitational instability (GI) of  accretion discs, in particular because of its vital role as a possible mechanism for planet formation at the outer parts of  protoplanetary discs (PPDs) and star formation in  AGNs \citep[for recent reviews about GI in the accretion discs see, e.g.,][]{rice16,kratter16}.  Gravitational stability of a disc with the surface density $\Sigma$, the sound speed $c_{\rm s}$, and the Keplerian angular velocity $\Omega$ is studied using    the Toomre parameter, i.e. $Q=c_{\rm s} \Omega / \pi G \Sigma $. If $Q$ drops to a value less than a threshold around unity, the disc becomes gravitationally unstable \citep{toomre64}. 

Early analytical models of self-gravitating disc are actually direct generalization of the \cite{shakura} model (hereafter; SS model) by considering non-Keplerian angular velocity of the disc and its thickness correction due to the self-gravity of the disc itself. In these studies, like the SS model, the turbulent viscosity coefficient  is simply an input parameter, and the momentum transfer mechanism is drastically simplified by the $\alpha$-prescription \citep[e.g.,][]{khajenabi2007}. However, it is important to keep in mind that  deviations from the Keplerian rotation and the thickness correction because of the self-gravity are both negligible so long as the total mass of the disc is much smaller than that of the central object. 

In other attempts to model self-gravitating discs, some authors proposed that one can close the set of the equations not by an energy equation, but by a self-regulation prescription related to the condition of the marginal gravitational stability of the disc \cite[e.g.,][]{bertin97, bertin99, bertin2001,lodato2001,sirko,Matzner2005}. These models where the gaseous disc is maintained in a state of marginal gravitational stability by keeping the Toomre parameter around its threshold value are also known as $Q-$disc. The reason for suppressing the energy equation is that some feed-back mechanisms are assumed to supply the necessary energy to prevent the Toomre parameter to fall to less than unity. In another type of $Q-$disc models, however, the energy equation is kept, but the angular momentum transport equation is suppressed by assuming that some additional momentum transport mechanisms exist in the disc to maintain the system in a marginally stable state \citep[][]{collin1999,collin2008}. We have to note that in these models, like in   the SS model, the viscosity parameter $\alpha$ is treated as an "assigned parameter". A different approach for modelling  a disc where its turbulence is driven by the GI, however, is to specify the viscosity parameter $\alpha$ as an explicit function of the Toomre parameter, but its exact functional form is prescribed in an ad hoc fashion \citep[e.g.,][]{Lin87,kratter2008,zhu2009,zhu2009b,zhu2010,martin2011,martin2013,martin2014}. In these models, not only the dependence of $\alpha$ on $Q$ is not determined based on physical arguments, but also it is not explicitly assumed that the Toomre parameter remains close to its critical value.

Even the Toomre parameter being very close to the threshold of the instability, however, does not imply that the disc will fragment within a reasonable period of time. It has been shown by \cite{gammie2001} and confirmed by subsequent studies \citep[e.g.,][]{Johnson,rice2003,rice2005,cossins2009,cossins2010}  that the fragmentation of a disc not only implies that the Toomre parameter being close to its critical value, but also it depends on the disc efficiency in losing the generated heat due to the gravity-driven turbulence. In other words, if the cooling timescale, $t_{\rm cool}$, becomes much longer than the dynamical timescale, $\Omega^{-1}$, even with a Toomre parameter close to its threshold, the disc will not fragment. Under these circumstances, instead, the disc can settle into a non-fragmentating  gravitoturbulent state where the angular momentum is transported by gravity-driven turbulence. It has been shown that the dimensionless stress parameter $\alpha$ is related to the cooling timescale as  $\alpha \simeq (\Omega t_{\rm cool})^{-1}$ \citep[e.g.,][]{gammie2001}. Numerical simulations show that when $\alpha \lesssim \alpha_{\rm C} \sim 1$, the disc fragments into a number of clumps and in the opposite limit, where the disc is in thermal equilibrium, the disc settles into a stationary {\it gravitoturbulent} state.  There are, however, uncertainties about the critical value $\alpha_{\rm C}$ \citep[e.g.,][]{rice2005}.

Thus, in a gravitoturbulent disc, the Toomre parameter $Q$ remains close to its critical value and the dimensionless stress parameter $\alpha$ is written in terms of the cooling timescale and the angular velocity. \cite{rafikov2009} explored  properties of the steady-sate gravitoturbulent discs where there are only two input parameter (i.e. central mass and the accretion rate)  and in contrast  to  most of  previous analytical models for the self-gravitating discs, the parameter $\alpha$ is no longer an input parameter. \cite{rafikov2009} studied  the astrophysical implications of his interesting analytical solutions in the context of PPDs. Although \cite{rafikov2009} properties of the gravitoturbulent discs in detail, \cite{Levin2007} had also proposed more or less a similar model for explaining star formation near to the Galactic center. None of these models included radiation pressure.

Recently, \cite{rafikov2015} extended his previous work by doing  a detailed comparative study between his solutions and those solutions which consider explicit dependence of the stress parameter $\alpha$ on the Toomre parameter  $Q$ in an ad hoc fashion \citep[e.g.,][]{Lin87,kratter2008,zhu2009,zhu2009b,zhu2010,martin2011,martin2013,martin2014}. He then showed that his approach is  more flexible, robust, and straightforward. Other properties of a PPD such as locations of  dead zone and  snow line are also investigated in \cite{rafikov2015}.

Although the concept of the gravitoturbulent state has been used for describing the outer parts of PPDs \citep[][]{rafikov2009,clarke2009,Matzner2005,rafikov2015}, the astrophysical implications of this model can  be extended to the self-gravitating quasar discs as well. Many authors have already investigated the  properties of the outer parts of the quasar discs which are prone to the gravitational instability using either analytical approaches \citep[e.g.,][]{shlo89,Levin2007,nayak2006} or numerical simulations \citep[e.g.,][]{goodman11} under certain simplifying assumptions.  Most of the analytical models for describing the inner regions of the  quasar discs, however,  are $Q-$disc models \citep[e.g.,][]{goodman2003,sirko,goodman2004,Levin2007} or only a direct generalization of SS model \citep[e.g.,][]{khajenabi2007} which means the $\alpha$ parameter is treated as an input parameter. These models predict that sub-parsec region of a quasar disc is gravitationally stable, but beyond a characteristic  radius where its location depends on the input parameters of the model, the disc  fragments into clumps which their subsequent evolution is under intense debate \citep[e.g.,][]{goodman2004,Levin2007,mckernan,Ina2015}. On a larger scale, however, these models are not appropriate and in more advanced models not only ongoing intense star formation, stars, and their gravitational interactions with the gas component should be considered, but also angular momentum transport  occurs by different mechanisms such as  global spiral waves, gravitational star-gas interactions and even supernova explosions  \citep[e.g.,][]{Thompson2005, wang2010,krum2010,  hopkins11,fur,Ina2015, inay2016}.

The advantage of gravitoturbulent model is that $\alpha$ parameter is obtained using physical arguments based on the thermodynamics of the disc. In fact, thermal equilibrium dictates a unique value for this parameter. In doing so, thermal content of the system and the rate of internal energy loss  are important physical gradients. A similar approach can also be adopted for modelling quasar discs, where both the gas and the radiation pressures are important. In the gravitoturbulent model of \cite{rafikov2009}, however,  radiation pressure is neglected because of its negligible role in the gravitational properties of a PPD. In addition to the dynamical role of radiation pressure by providing extra support in the vertical direction of a disc, radiation pressure also modifies the thermal energy content of a disc and the viscosity parameter. 

A gravitoturbulent model is presented in this study which is appropriate for the self-gravitating  inner part of a quasars disc. Our analytical solutions are different from the \cite{rafikov2009} solutions in that our model  includes radiation pressure as well. In the next section, basic assumptions and equations are presented. Optically thick solutions are obtained in section 3. Properties of the solutions, location of the fragmentation boundary, and the mass of the clumps are studied for a wide range of the input parameters in section 4.  We  show that if a fixed value for the critical stress parameter $\alpha_{\rm C}$ is considered, in the presence of radiation and for high accretion rates, the radius beyond which the disc fragments is  unphysically large. Overcoming this problem requires $\alpha_{\rm C}$ to be an increasing power-law function of the ratio of gas and total pressures which is actually consistent with numerical simulations \citep[e.g.,][]{goodman11}. In section 5, an estimate of the central black hole (BH) maximum mass using our solution is presented. We then conclude by a summary of the results and possible astrophysical implications in the last section.

\section{Basic Equations}
We consider a gravitoturbulent disc in which not only the gas pressure $p_{\rm gas}$ is important, but also the radiation pressure $p_{\rm rad}$ plays a significant  role. Thus, we can introduce total pressure as $p = p_{\rm gas} + p_{\rm rad}$, where $p_{\rm gas} = \rho k_{\rm B} T / \mu m_{\rm H}$ and in the {\it optically thick regime} the radiation pressure becomes $p_{\rm rad} = \frac{1}{3} a T^4$. The density and the midplane temperature are denoted by $\rho$ and $T$, and  $\mu$ is the mean molecular weight and $m_{\rm H}$ is the mass of Hydrogen. We adopt $\mu =0.6$ that is valid for a fully-ionized gas. Moreover, $k_{\rm B}$ is Boltzmann constant and $a$ is the radiation constant. Then, it will be useful to introduce the ratio of the gas pressure and the total pressure as $\beta = p_{\rm gas} / p$. We then obtain
\begin{equation}\label{eq:temp-1}
\frac{1}{3}a \left (\frac{\mu m_{\rm H}}{k_{\rm B}} \right )\frac{T^3}{\rho}=\frac{1-\beta}{\beta}. 
\end{equation}

We also assume that the disc is in a self-regulated state which means  that its Toomre parameter stays around threshold value of the instability, say $Q_0$, and so $Q_0 = \Omega c_{\rm s}/ (\pi G \Sigma )$.  On the other hand, the surface density$\Sigma$ is written in terms of the volume density $\rho$ and the thickness of the disc $H$, i.e. $\Sigma = 2\rho H$, where $H=c_{\rm s}/\Omega $, $\Omega = (GM/r^3)$ and $c_{\rm s}=\sqrt{p/\rho} $. Here, the radial distance is denoted by $r$. Upon substituting these relations into the above self-regulated condition, the volume density is obtained as 
\begin{equation}\label{eq:density}
\rho =\frac{\Omega^2}{2\pi G Q_0}.
\end{equation}
Making use of the Equations (\ref{eq:temp-1}) and (\ref{eq:density}), the midplane temperature can now be written as
\begin{displaymath}
T = \left (\frac{2}{3}   \right )^{-1/3} (\pi  G Q_0)^{-1/3} a^{-1/3 }\left (\frac{\mu m_{\rm H}}{k_{\rm B}} \right )^{-1/3}
\end{displaymath}
\begin{equation}\label{eq:temp-f}
\times \left (\frac{1-\beta}{\beta} \right )^{1/3} \Omega^{2/3}.
\end{equation}

From the self-regulated condition, the sound speed is obtained as $c_{\rm s} = \pi G Q_0 \Sigma /\Omega $.  If we substitute from equation (\ref{eq:density}) for the density, a relation between the total pressure and the surface density is obtained, i.e.
\begin{equation}\label{eq:p}
p = \frac{1}{2} (\pi G Q_0 ) \Sigma^2. 
\end{equation}
On the other hand, the total pressure is $p=\rho k_{\rm B} T / \mu m_{\rm H} + (1/3) a T^4$ and upon substituting into this equation from equation (\ref{eq:temp-f}) for the temperature and from equation (\ref{eq:p}) for the total pressure, the surface density is obtained in terms of the ratio $\beta$ and angular velocity, i.e.
\begin{displaymath}
\Sigma = \left (\frac{2a}{3} \right )^{-1/6} \left (\pi G Q_0 \right )^{-7/6} \left (\frac{\mu m_{\rm H}}{k_{\rm B}} \right )^{-2/3} 
\end{displaymath}
\begin{equation}\label{eq:surf}
\times \Omega^{4/3} \left (1-\beta \right )^{1/6} \beta^{-2/3}.
\end{equation}

Under the assumption of complete ionization and equal gas and radiation temperatures, the internal energy per unit area, $U$, can be written as \citep{goodman11},
\begin{equation}
U = (1-\frac{\beta}{2}) \Sigma c_{\rm s}^2 .
\end{equation}
The cooling function $\Lambda$ describes the radiative losses  from the surface of the disc, i.e.
\begin{equation}
\Lambda = 2\sigma T_{\rm eff}^4 \approx \frac{16}{3} \sigma T^4 f(\tau )^{-1}
\end{equation}
where $\sigma$ is the Stephan-Boltzman constant, $T_{\rm eff}$ is the effective temperature at the surface of the disc and $\tau$ is the optical depth. Here, function $f(\tau )$ is introduced to smoothly interpolates between optically thick ($\tau \gg 1$) and optically thin ($\tau \ll 1$) regimes. This function is approximated by  $f(\tau ) = \tau + \tau^{-1}$. Having the internal energy and the cooling function, the cooling time, $t_{\rm cool}$, becomes
\begin{equation}
t_{\rm cool} = \frac{U}{\Lambda} = \frac{(1-\frac{\beta}{2})\Sigma c_{\rm s}^2}{\frac{16}{3} \sigma T^4} f(\tau ).
\end{equation}

 Now, we can obtain the dimensionless stress parameter  $\alpha$  as  a function of $\Omega$ and $\beta$. Upon substituting the cooling time into equation $\alpha \simeq (\Omega t_{\rm cool})^{-1}$, we have
\begin{equation}\label{eq:alpham}
\alpha = \frac{16}{3} \sigma T^{4} (1-\frac{\beta}{2})^{-1} \Sigma^{-1} \Omega^{-1} c_{\rm s}^{-2} [f(\tau )]^{-1}.
\end{equation}
 Since in a gravitoturbulent disc the Toomre parameter is about its threshold for the instability, we then obtain $c_{\rm s}= \pi G Q_0 \Sigma /\Omega $. Using this equation for the sound speed, equation (\ref{eq:alpham}) is written as 
\begin{equation}
\alpha = \frac{16}{3} \sigma T^4 (\pi G Q_0 )^{-2} (1-\frac{\beta}{2})^{-1} \Sigma^{-3} \Omega [f(\tau )]^{-1}.
\end{equation}
 If we use equation (\ref{eq:surf}) for the surface density, the above equation becomes
\begin{displaymath}
\alpha = 8 (\frac{2}{3})^{1/6} \sigma \left( \pi G Q_0 \right )^{1/6} a^{-5/6} \left ( \frac{\mu m_{\rm H}}{k_{\rm B}}\right )^{2/3}
\end{displaymath}
\begin{equation}\label{eq:alphaf}
\times \Omega^{-1/3} (1-\frac{\beta}{2})^{-1} (1-\beta )^{5/6} \beta^{2/3} [f(\tau )]^{-1}.
\end{equation}
 Also, viscosity is $\nu = \alpha c_{\rm s}^2 /\Omega$, or,
\begin{equation}
\nu = \alpha (\pi G Q_0 )^2 \Sigma^{2} \Omega^{-3}.
\end{equation}
 By substituting equations (\ref{eq:surf}) and (\ref{eq:alphaf}) into the above equation for the viscosity, we obtain
\begin{displaymath}
\nu = 8 \left (\frac{2}{3} \right )^{-1/6} \sigma \left( \pi G Q_0 \right )^{-1/6} a^{-7/6} \left ( \frac{\mu m_{\rm H}}{k_{\rm B}}\right )^{-2/3}
\end{displaymath}
\begin{equation}\label{eq:nu}
\times  \Omega^{-2/3} (1-\frac{\beta}{2})^{-1} (1-\beta )^{7/6} \beta^{-2/3} [f(\tau )]^{-1}.
\end{equation}

 So far we have obtained physical quantities of the disc as functions of $\Omega$ and $\beta$, and now, another relation between these variables is needed to close the equations of our model. We know that $\dot{M}=3\pi \nu \Sigma$, where $\dot{M}$ is the accretion rate. Upon substituting from equations (\ref{eq:surf}) and (\ref{eq:nu}) into this equation, an algebraic equation is obtained, i.e., 
\begin{displaymath}
\dot{M}= 24\pi \left (\frac{2}{3} \right )^{-1/2} \sigma  \left( \pi G Q_0 \right )^{-4/3}  a^{-4/3} \left ( \frac{\mu m_{\rm H}}{k_{\rm B}}\right )^{-4/3}
\end{displaymath}
\begin{equation}\label{eq:rate}
\times  \Omega^{2/3} (1-\frac{\beta}{2})^{-1} (1-\beta )^{4/3} \beta^{-4/3} [f(\tau )]^{-1}.
\end{equation}

The above equations describe steady-state structure of a gravitoturbulent disc including radiation and gas pressures. When the system is optically thick, we can further simplify these solutions and  transform them into dimensionless as we do in the next section.

\section{Optically thick solutions}
We can now consider optically thick regime where the optical depth is much larger than one. Thus, we have  $\tau = \frac{1}{2} \kappa_{\rm e.s.} \Sigma \gg 1$, where $\kappa_{\rm e.s.} \approx 0.4$ cm$^2$ g$^{-1}$ is the electron-scattering opacity. In the optically thick regime, we have $f(\tau ) \simeq \tau $. Dimensionless accretion rate $\dot{m}$, central mass $m$ and the radial distance $\tilde{r}$ are introduced as
\begin{equation}
\dot{m} = \frac{\dot{M}}{\dot{M}_0},  m= \frac{M}{M_0}, \tilde{r} = \frac{r}{r_0}, 
\end{equation}
where $\dot{M}_0 $, $M_0$ and $r_0$ are the reference values for the accretion rate, mass and the distance, respectively. Using Eddington luminosity,  $L_{\rm Edd}$, the Eddington accretion rate is defined as $\dot{M}_{\rm Edd} = L_{\rm Edd}/\epsilon c^2 = 4\pi G M / \epsilon \kappa_{e.s.} c$, where $c$ is the speed of light and $\epsilon$ is the accretion efficiency which depends on the BH spin \citep{bardeen}. A value that is often used is $\epsilon=0.1$ and we adopt this value unless otherwise stated. As for the reference of the accretion rate, we assume $\dot{M}_0 = \dot{M}_{\rm Edd}$. We also assume $r_{0}=r_{\rm s}$, where $r_{\rm s}$ is the Schwarzschild radius, i.e. $r_{\rm s}=2GM/c^2$.  Thus, equation (\ref{eq:rate}) becomes
\begin{equation}\label{eq:rtild}
\tilde{r}(\beta ) = \dot{m} m^{1/3} C^{-1} (1-\frac{\beta}{2}) (1-\beta )^{-7/6} \beta^{2/3}
\end{equation}
where the dimensionless parameter $C$ is
\begin{displaymath}
C = 48\pi \left (\frac{2}{3} \right )^{-1/3} \sigma \kappa_{\rm e.s.}^{-1} \left( \pi G Q_0 \right )^{-1/6}  a^{-7/6} 
\end{displaymath}
\begin{equation}
\times \left ( \frac{\mu m_{\rm H}}{k_{\rm B}}\right )^{-2/3} \dot{M}_{0}^{-1} \Omega_{0}^{-2/3},
\end{equation}
where $\Omega_0 = (GM_0 / r_{0}^3 )^{1/2}$. The surface density becomes
\begin{equation}\label{eq:sigma}
\Sigma (\beta ) = \Sigma_{0} C^2 \dot{m}^{-2} (1-\frac{\beta}{2})^{-2} (1-\beta )^{5/2} \beta^{-2},
\end{equation}
where
\begin{equation}
\Sigma_{0} = \left (\frac{2}{3} \right)^{-1/6}  a^{-1/6} \left( \pi G Q_0 \right )^{-7/6} \left (\frac{\mu m_{\rm H}}{k_{\rm B}} \right )^{-2/3} \Omega_{0}^{4/3}.
\end{equation}
The stress parameter becomes
\begin{equation}
\alpha (\beta ) = \alpha_0 C^{-1/2} \dot{m}^{5/2} (1-\frac{\beta}{2})^{3/2} (1-\beta )^{-9/4} \beta^3,
\end{equation}
\begin{displaymath}
\alpha_0 = 16 \left(\frac{2}{3} \right)^{1/6} \sigma \left( \pi G Q_0 \right )^{1/6} a^{-5/6} \left (\frac{\mu m_{\rm H}}{k_{\rm B}} \right )^{2/3} 
\end{displaymath}
\begin{displaymath}
\times \Omega_{0}^{-1/3} \kappa_{\rm e.s.}^{-1} \Sigma_{0}^{-1} C^{-2}.
\end{displaymath}
Also, temperature of the disc is obtained as
\begin{equation}
T=T_0 \dot{m}^{-1} (1-\frac{\beta}{2})^{-1} (1-\beta )^{3/2} \beta^{-1},
\end{equation}
\begin{equation}
T_0 = \left (\frac{2}{3}   \right )^{-1/3}  (\pi  G Q_0)^{-1/3} a^{-1/3 }\left (\frac{\mu m_{\rm H}}{k_{\rm B}} \right )^{-1/3}  \Omega_{0}^{2/3} C.
\end{equation}
 We can also determine the thickness of the disc, i.e. $H=c_{\rm s}/\Omega$. Using self-regulated condition, thickness of the disc becomes $H=(\pi G Q_0 ) \Sigma/\Omega^2$. Upon substituting from equations (\ref{eq:rtild}) and (\ref{eq:sigma}) into this equation, we can obtain the opening angle of the disc, i.e.,
\begin{equation}
\frac{H}{r}= H_0 (1-\beta )^{29/6} \beta^{-2/3},
\end{equation} 
where $H_0 = (\pi G Q_0 ) (GM)^{-1} r_{0}^2 \Sigma_0 m^{2/3}$. This equation shows that when the radiation pressure is high (i.e., low $\beta$ regime), the ratio $H/r$ is larger compared to a case with a low radiation pressure. It means that the disc becomes slim rather than thin. 

The above analytical solutions describe properties of an optically thick disc  with radiation. In the next section, we explore these solutions for different sets of the input parameters.
\begin{figure*}
\includegraphics[scale=1.0]{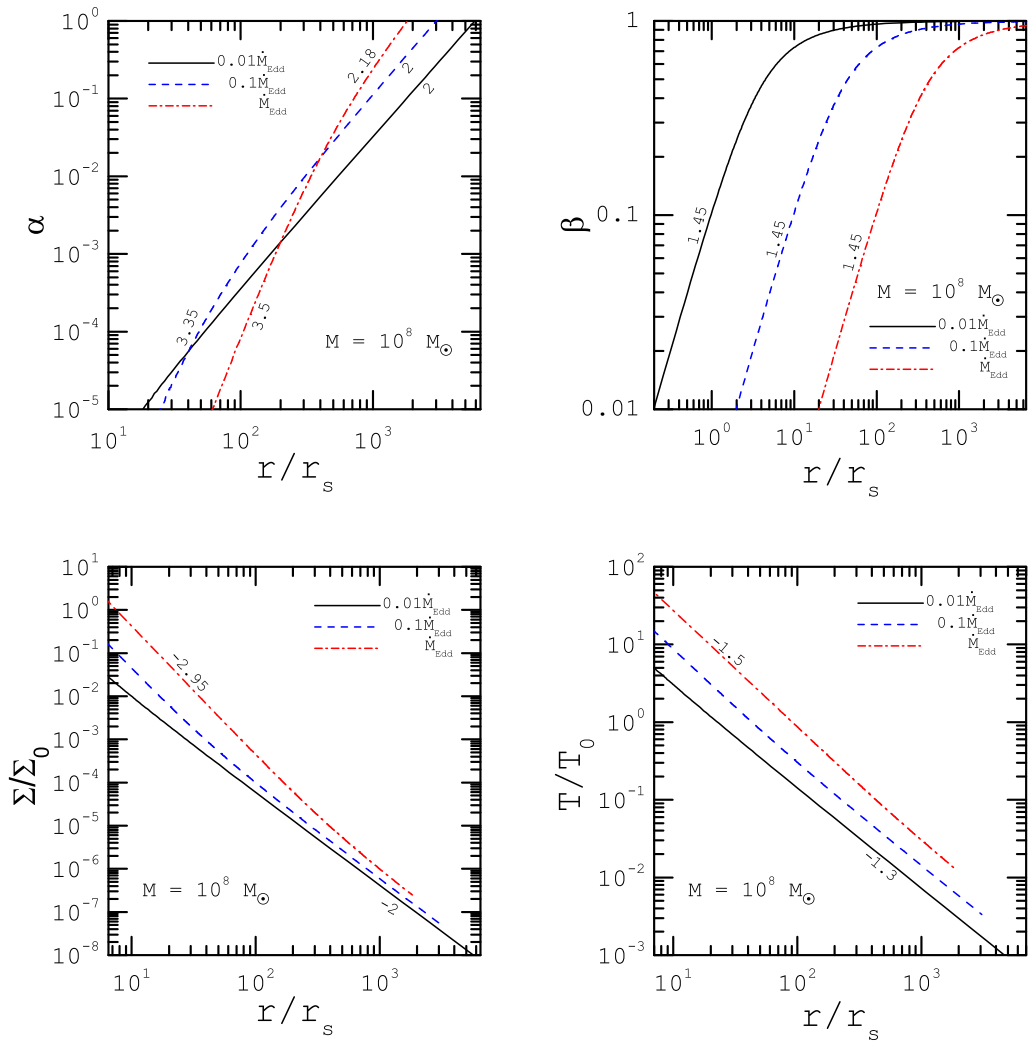}
\caption{Profiles of the dimensionless stress paramerer (top, left), ratio of the gas pressure and the total pressure (top, right), surface density (bottom, left) and temperature (bottom, right) versus radial distance normalized by the Schwarzschild radius. Mass of the central black hole is $10^8$ ${\rm M}_{\odot}$ and $\mu = 0.6$. Here, we have $\Sigma_{0}=2.34\times 10^{11}$ g cm$^{-2}$ and $T_0 = 9.2\times 10^4 $ K. These curves are shown for different accretion rates: $\dot{m}=0.01$ (solid), $0.1$ (dashed) and $1$ (dashed-dot). For each physical quantity, we found that it is possible to fit a power-law function of the radial distance and the corresponding exponent is shown near to the curve.}\label{fig:f1}
\end{figure*}

\begin{figure*}
\includegraphics[scale=1.0]{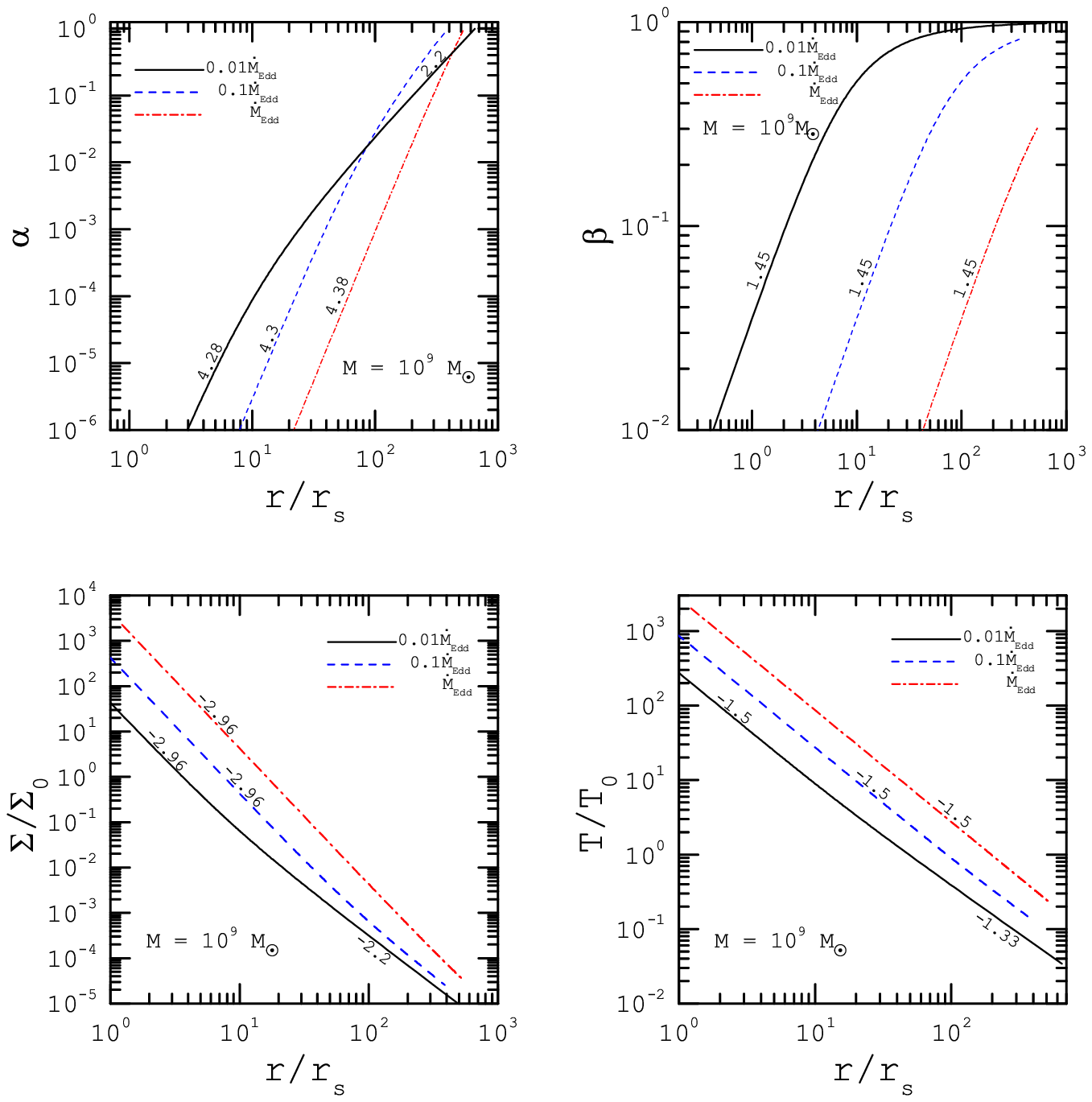}
\caption{Same as  Figure \ref{fig:f1}, but the mass of the central black hole is $M=10^9$ ${\rm M}_{\odot}$. Here, we have $C=2.37\times 10^{-3}$, $\Sigma_{0} = 2.34\times 10^{9}$ g cm$^{-2}$, and $T_0 = 9.2\times 10^3$ K.}\label{fig:f2}
\end{figure*}
\section{Analysis}
\subsection{Properties of the solutions}
Given the mass of the central black hole and the rate at which  mass is accreting onto  it, we can  explore the behavior of our solutions as a function of the radial distance. Figure \ref{fig:f1} shows stress parameter $\alpha$ (top, left), ratio $\beta$ (top, right), surface density (bottom, left) and the temperature profiles (bottom, right) of a disc with central mass $10^8$ solar masses. The role of the second input parameter, i.e. accretion rate, is illustrated by considering different accretion rates: $\dot{M}=0.01 \dot{M}_{\rm Edd}$ (solid), $0.1 M_{\rm Edd}$ (dashed) and $M_{\rm Edd}$ (dashed-dot). Our solutions exhibit a strong dependence on accretion rate. For a central BH with  mass  $10^8$ M$_{\odot}$, we then obtain $C=2.37\times 10^{-3}$, $\Sigma_{0} = 2.34\times 10^{11}$ g cm$^{-2}$, $\alpha_0 = 3.11\times 10^{-3}$, $T_0 = 9.2\times 10^4$ K. Number beside each curve is the exponent of a  power-law function of the radial distance that can be fitted to the shown curve. We find that these approximate power-law functions describe our solutions reasonably well, i.e. 
\begin{equation}
\alpha \propto r^{\nu_{\alpha}}, \beta \propto r^{\nu_{\beta}}, \Sigma \propto r^{\nu_{\Sigma}}, T \propto r^{\nu_T},
\end{equation}
where the slops $\nu_{\alpha}$, $\nu_{\beta}$, $\nu_{\Sigma}$ and $\nu_T$ strongly depend on the input parameters. All shown solutions give optical depths much larger than one as we expect in the optically thick regime.

The profile of $\alpha$ versus radial distance for high accretion rates tends to be steeper at small radii comparing to the outer parts. At the inner parts the slope is between $\nu_{\alpha}\simeq 3.35$ and $3.5$,  however, all curves tend to a slope around 2 at the outer regions of the disc. The transition between these two regions shifts to the smaller radii as the accretion rate decreases. As we go to a higher accretion rate, the stress parameter $\alpha$ gets closer to its critical value at a smaller radius, which means the disc becomes increasingly susceptible to the fragmentation with increasing the accretion rate.   We deem that a disc fragments once $\alpha$ becomes larger than its critical value $\alpha_{\rm C} \simeq 1$, and so all solutions are truncated at a charachteristic  radius where $\alpha = \alpha_{\rm C}=1$. In the absence of radiation, \cite{rafikov2009} found that $\alpha \propto r^{4.5}$ irrespective of the accretion rate. This suggests that radiation causes  to have a shallower profile  of $\alpha$ and its slope depends on the accretion rate. Since variation of $\alpha$ is slower than  \cite{rafikov2009}, the disc evolves as a gravitoturbulent state over a larger range of radii  compared to a case without radiation.

The profile of $\beta$ as a function of $r$ is shown in Figure \ref{fig:f1} (top, right). While the disc is radiation-dominated at its inner regions, as we go to the outer parts, role of  radiation becomes less effective so that $\beta$ tends to one at the self-gravitating radius\footnote{We define self-gravitating radius as a radius where the stress tensor parameter reaches to its critical value and so, the disc may fragment into clumps.}. This trend is more or less independent of the accretion rate for the explored cases in this figure, however, the action of the radiation is more effective over a wider range of the radial distances for high accretion rates. Moreover, the slope is $\nu_{\beta} \simeq 1.45$, irrespective of the accretion rate.

Figure \ref{fig:f1} (bottom, left) depicts surface density versus $r$  and demonstrates that its slope varies between $-2$ to $-3$ depending on the accretion rate. Surface density at the inner part of the disc with a slope around -3 is steeper than the outer part  with a slope -2, and, the transition between these two regions shifts to the larger radii with increasing the accretion rate. For sub-Eddington accretion rates,  over  a large spatial extend of the disc, the slope of the surface density distribution is -2 which agrees with a trend found by \cite{goodman11} in their  simulations. With increasing the accretion rate, not only the surface density enhances at all parts of the disc, but also the size of inner part with a slope -3 becomes larger, though the slope of the outer part remains around -2. The temperature of the disc also exhibits a strong dependence on the accretion rate as we can see in Figure \ref{fig:f1} (bottom, right). The disc is indeed cooler in the outer parts for a given accretion rate, and, as expected, the disc is hotter with increasing the accretion rate. The slope of temperature, however, shows little variations with the accretion rate. As we go to higher  accretion rates, the temperature profile becomes slightly steeper so that its slope varies from about -1.3 to -1.5.    

\begin{figure}
\includegraphics[scale=0.5]{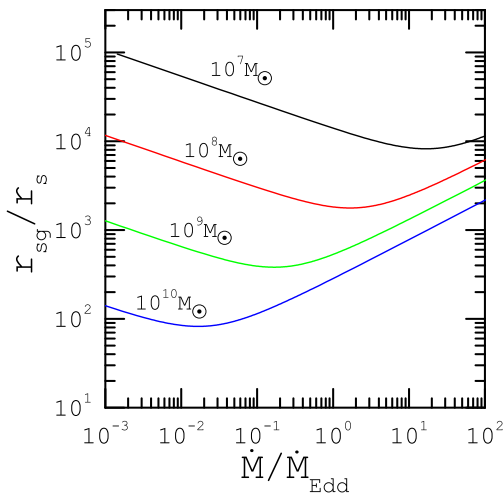}
\caption{Dependence of the self-gravitating radius $r_{\rm sg}$ on the accretion rate. Each curve is labeled by the corresponding mass of the central black hole in solar masses.}\label{fig:f3}
\end{figure}

\begin{figure}
\includegraphics[scale=0.5]{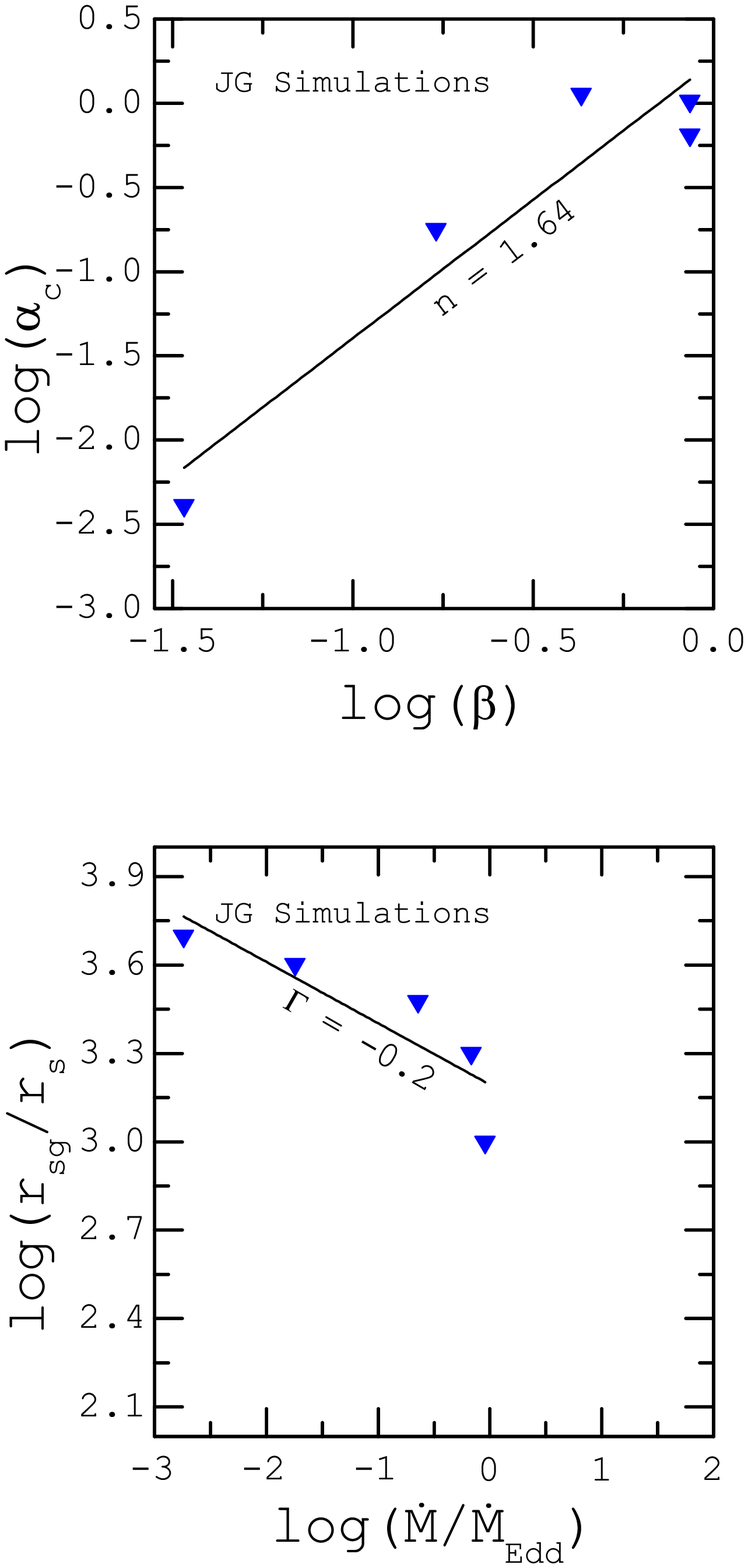}
\caption{Results from numerical simulations are marked with  triangles.}\label{fig:f4}
\end{figure}

\begin{figure}
\includegraphics[scale=0.5]{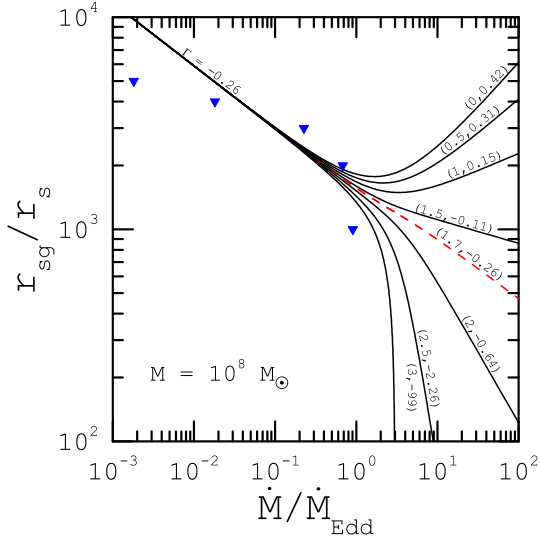}
\caption{Self-gravitating radius $r_{\rm sg}$ as a function of the accretion rate. The mass of the central black hole is $10^8$ ${\rm M}_{\odot}$. Each curve is labeld by a pair of numbers as $(n,\Gamma)$, where $n$ is the slope of the critical stress parameter and $\Gamma$ is the slope of the fitted function as $r_{\rm sg} \propto \dot{m}^{\Gamma}$. A case with $n=1.7$ which is the based on the results of the simulations (solid triangles) is shown by dashed curve. }\label{fig:f5}
\end{figure}

\begin{figure}
\includegraphics[scale=0.5]{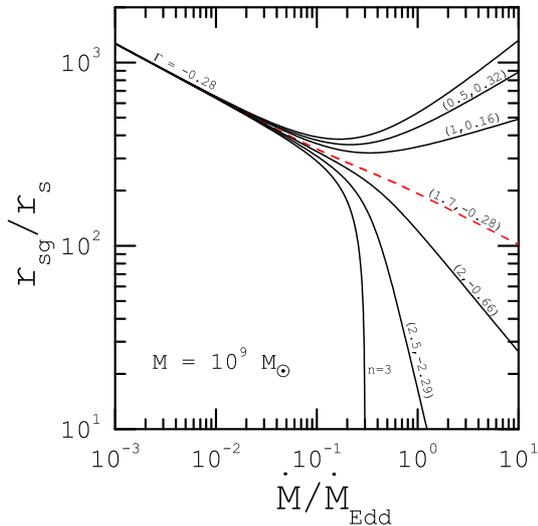}
\caption{Same as Figure \ref{fig:f5}, but for a more massive central object, i.e. $M=10^9$ M$_{\odot}$.}\label{fig:f6}
\end{figure}

Now, we explore how these results are influenced by varying the mass of the central object. In Figure \ref{fig:f2}, we plot disc quantities  for the same parameters as in Figure \ref{fig:f1}, but for a central BH with  $M=10^9$ $M_{\odot}$. 	In this case, we have $C=2.37\times 10^{-3}$, $\Sigma_{0} = 2.34\times 10^{9}$ g cm$^{-2}$, $\alpha_0 = 0.98$, $T_0 = 9.2\times 10^3$ K.	
The profiles of $\alpha$ for different accretion rates are shown in \ref{fig:f2} (top, left). This plot shows  that $\alpha (r)$ is very steep in the inner parts of the disc with a slope $\nu_\alpha \simeq 4.3$, whereas slope of about 2.2 is achieved further out in the disc. As the accretion rate increases, however, the slope $\nu_\alpha$ tends to about 4.38 over entire range of the radial distances.

The behavior of the ratio $\beta$ is shown in Figure \ref{fig:f2} (top, right). While the inner parts are radiation-dominated independent of the accretion rate, the outer regions are gas-dominated for sub-Eddington or Eddington accretion rates.  As before, all curves are truncated  once the stress parameter $\alpha$ becomes one. Profile of the surface density (bottom, left) and temperature (bottom, right) show that their slopes are $\nu_{\Sigma} \simeq -3$ and $\nu_{T} \simeq -1.5$ for Eddington accretion rate, whereas for sub-Eddington accretion rates, the profiles become shallower with  slopes about $\nu_{\Sigma} \simeq -2.2$ and $\nu_{T} \simeq -1.33$. Thus, the general trends of the physical quantities of a disc with $M=10^9$ M$_{\odot}$ are similar to the case  with  $M=10^8$ M$_{\odot}$ in Figure \ref{fig:f1}.

 Although we introduced function $f(\tau )$ which provides a smooth transition between  optically thin and thick regimes, we found that all the solutions are actually corresponding to the optical thick case. So, we simplified the equations further in section 3 by writing $f(\tau )\simeq \tau$. Then, we determined physical quantities including temperature as a function of the radial distance. It means that the transition between low and high T regimes occurs in the optical thick regime, and the bridging formula $f(\tau) $ does not affect our solutions.  In section 3, we kept $f(\tau)$ only to present the equations in their general forms.

\subsection{Location of the self-gravitating radius}
In keeping with the approach outlined in the previous studies \citep[e.g.,][]{rafikov2009,rafikov2015}, the disc fragments into small clumps when stress parameter $\alpha$ drops below a critical value $\alpha_{\rm C}$. Figure \ref{fig:f3} shows the self-gravitating radius $r_{\rm sg}$ as a function of the accretion rate for various central masses. The expected trend is to have a smaller self-gravitating radius as the accretion rate increases. Contrary to this physical expected behavior, however, we see in Figure \ref{fig:f3} that for a given central mass, the radius $r_{\rm sg}$ decreases with $\dot{M}$, but beyond a certain accretion rate this behavior is reversed and radius $r_{\rm sg}$ increases as more mass is accreted. Note that for determining $r_{\rm  sg}$ in this figure, a fixed value for the critical value (independent of the properties of disc) is considered, i.e. $\alpha_{\rm C} =1$.  This unphysical trend for the self-gravitating radius as a function of the accretion rate, however, is significant for high accretion rates.

Figure \ref{fig:f3} raises some concerns about the validity of $\alpha_{\rm C} \simeq 1$ in the discs with the radiation pressure. Numerical simulations of \cite{goodman11} have already shown that once radiation pressure is included, fragmentation can occur at the boundary with $\Omega t_{\rm cool} \gg 1$, or equivalently $\alpha_{\rm C} \ll 1$. In other words, the critical value of the stress parameter $\alpha_{\rm C}$ depends on the ratio of the gas pressure and the total pressure, i.e. $\beta$. Numerical simulations of \cite{goodman11} are restricted to only a central black hole with $10^8$ solar masses, however, their results clearly show that $\alpha_{\rm C}$ depends on the ratio $\beta$. In Figure \ref{fig:f4} (top), we take values of the critical stress parameter $\alpha_{\rm C}$ directly from their plots, when available.  This Figure displays $\alpha_{\rm C}$ as a function of $\beta$ by solid triangles. We then found that a power-law function as $\alpha_{\rm C} \propto \beta^{n}$ can be fitted, where its  slope is about $n=1.64$. Bottom plot of Figure \ref{fig:f4} shows self-gravitating radius as a function of the accretion rate based on the simulations of \cite{goodman11}. Most noteworthy is the fact that our model for a disc with $10^8$ solar masses as a central object predicts that  location of $r_{\rm sg}$ is in the gas-dominated regime (i.e., $\beta \approx 1$) when the accretion rate is less than Eddington rate. Simulations of \cite{goodman11} also confirms this trend, whereas \cite{goodman2004} predict that location of $r_{\rm sg}$ is in the radiation-dominated part of the disc. Here, we fitted a power-law function as $r_{\rm sg}/r_{\rm s} \propto \dot{m}^{\Gamma}$, where its slope is around $\Gamma = -0.2$.

Thus, the critical parameter $\alpha_{\rm C}$ turns out to depend  on the ratio $\beta$ as a power-law function. Neglecting this dependence leads to unexplainable trends for the fragmentation boundary, as we showed in Figure \ref{fig:f3}. Rather than adopting a fixed value around unity for the critical stress parameter as in \cite{rafikov2009} and some other previous works, we instead use $\alpha_{\rm C}=\beta^n$, where in the absence of radiation it tends to unity, and, this new condition will enable us to determine $r_{\rm sg}$ self-consistently. 

Figure \ref{fig:f5} illustrates the dependence of the self-gravitating radius on the accretion rate for a central black hole with $10^8$ solar masses and various slope $n$. Each curve is labeled by a pair of numbers, where the first number is $n$ and the second number denotes the slope $\Gamma$. A case with $n=1.7$  which agrees with the simulations is shown by the dashed curve. Results of the simulations are marked by solid triangles. For sub-Eddington accretion rates, location of $r_{\rm sg}$ is independent of the slope $n$ and its dependence on the accretion rate can be fitted as $r_{\rm sg} \propto \dot{m}^{-0.26} r_{\rm s}$. This result agrees reasonably well with the simulations of \cite{goodman11} which can be fitted as $r_{\rm sg} \propto \dot{m}^{-0.2} r_{\rm s}$ (see Figure \ref{fig:f4}). Our analysis for the Eddington accretion rate predicts that $r_{\rm sg} \simeq 1600 r_{\rm s}$, whereas fitted function to the simulations gives $r_{\rm sg} \simeq 2570 r_{\rm s}$. But equation (39) in \cite{goodman11} for a similar accreting system gives $r_{\rm sg} \simeq 4000 r_{\rm s}$. In a disc with a fixed stress parameter and viscous tensor in proportion to the total pressure, on the other hand, \cite{goodman2003} found that the self-gravitating radius becomes $r_{\rm sg}\simeq 2200 r_{\rm s}$, if $\beta \ll 1$, $M=10^8$ M$_{\odot}$ and $\dot{M}=\dot{M}_{\rm Edd}$. Once the accretion rate increase to the values larger than the Eddington rate, however, the effect of $n$ becomes more significant. For high accretion rates, fragmentation boundary resides  in the outer parts of the disc where the parameter $\beta$ is less than unity which means the effect of radiation pressure is more noticeable. For lower accretion rates, we have roughly $\beta (r_{\rm sg})\sim 1$ which implies a gas-dominated region. Figure \ref{fig:f6} is same as Figure \ref{fig:f5}, but for a central black hole with $10^9$ solar masses. Behavior of $r_{\rm sg}$ as a function of the accretion rate is analogous to the previous explored case with $M=10^8$ M$_{\odot}$, except for the fact that the effect of $n$ becomes noticeable at lower accretion rates compared to Figure \ref{fig:f5}. Again, the preferred value for $n$ turns out to be $1.7$ in agreement with the mentioned simulations.

Having explored different cases, we found an approximate relation for the self-gravitating radius as a function of the accretion rate and the central mass:
\begin{equation}\label{eq:rsg}
\frac{r_{\rm sg}}{r_{\rm s}} \simeq 1.57 \times 10^3 m^{-0.94} \dot{m}^{-0.28}.
\end{equation}

\subsection{Mass of the clumps}
\begin{figure}
\includegraphics[scale=0.65]{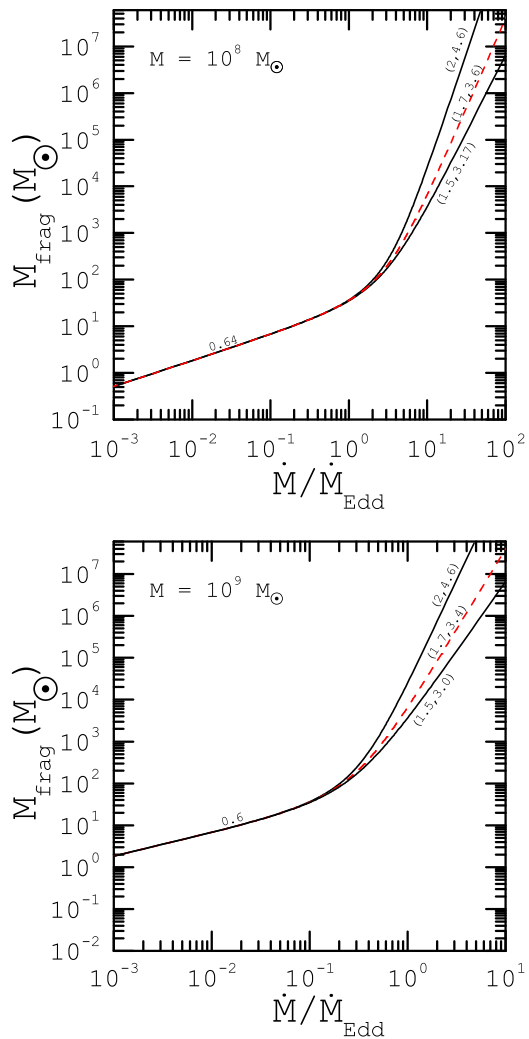}
\caption{Mass of the fragments at the self-gravitating radius, $M_{\rm frag}$, as a function of the accretion rate for two accreting systems with $M=10^8$ M$_{\odot}$ (top) and $M=10^9$ M$_{\odot}$ (bottom). We found that $M_{\rm frag}$ can be approximated as a power-law function of the accretion rate, i.e. $M_{\rm frag} \propto \dot{m}^\xi$, where the slope $\xi$ depends on the accretion rate. For the low accretion rate we have $\xi \simeq 0.6$ and this slope is independent of $n$, however, for higher accretion rates, each curve is labeled by a pair of numbers as $(n, \xi)$. A case corresponding  to $n=1.7$ is shown by a dashed curve.}\label{fig:f7}
\end{figure}

Now, we can estimate mass of the fragments at the self-gravitating radius. The most unstable mode for a marginally gravitationally stable disc is of order of the disc thickness, $H$. Thus, the mass of a fragment at the self-gravitating radius becomes $M_{\rm frag}\approx 4\pi \Sigma H^2$. Using our solutions, we then obtain
\begin{displaymath}
M_{\rm frag} = 4 \pi^2 \left( \pi G Q_0 \right)^{-3/2} \left(\frac{2a}{3} \right)^{-1/2} \left(\frac{\mu m_{\rm H}}{k_{\rm B}} \right)^{-2}
\end{displaymath}
\begin{equation}
\times \frac{\left(1-\beta_{\rm sg} \right)^{1/2}}{\beta_{\rm sg}^2},
\end{equation}
where $\beta_{\rm sg}$ represents value of this ratio at the self-gravitating radius. Obviously, as the ratio $\beta_{\rm sg}$ tends to unity which corresponds to a gas-dominated region at the fragmentation boundary, the mass of the fragments, $M_{\rm frag}$, decreases. But if radiation becomes significant at the self-gravitating radius which means $\beta_{\rm sg}$ tends to zero, the mass $M_{\rm frag}$ becomes very large. Figure \ref{fig:f7} shows mass of the fragments as a function of the accretion rate for different central masses, i.e. $M=10^8$ M$_{\odot}$ (top) and $M=10^9$ M$_{\odot}$ (bottom). For finding location of the self-gravitating radius, different values of the slope $n$ are considered, however, both the radius $r_{\rm sg}$ and the mass $M_{\rm frag}$ are independent of the slope $n$ for low accretion rates. For high accretion rates, on the other hand, the mass $M_{\rm frag} $ depends on both the accretion rate and the slope $n$. The profiles of Figure \ref{fig:f7} can be approximated as a function like $M_{\rm frag} \propto \dot{m}^\xi$. We found that $\xi\simeq 0.6$ for low accretion rates. Curves corresponding to the high accretion rates are labeled by a pair of numbers, i.e. $(n, \xi)$. Note that although the  fragmentation boundary and the mass of the clumps at this radius are estimated, we can treat neither details of fragmentation process nor the subsequent dynamical evolution of clumps.

\section{Estimate of the maximum black hole mass}
Observations have revealed that the mass of the SMBHs is between $10^5$ and $10^{10}$ M$_{\odot}$. Is there any limit on the maximum mass of a SMBH? Theoretical attempts to address this question started by \cite{natar2009} and then improved in some aspects by \cite{king2016} and \cite{inay2016}. Most of the previous models for SMBH accretion disc, as we discussed earlier, predicted that the disc extend up to the self-gravitating radius and  beyond this radial distance, the disc is susceptible  to gravitational instability and it may fragment into clumps. Thus, the outer radius of a SMBH disc cannot be larger than $r_{\rm sg}$. Our gravitoturbulent model confirms this finding, however, the precise value of the self-gravitating radius depends on the input parameters, namely, mass of the central BH and the accretion rate.   But the radius  $r_{\rm sg}$ should exceed the innermost stable circular orbit $r_{\rm ISCO}$, and if not; the disc accretion is suppressed. \cite{king2016} estimated the maximum mass of a SMBH following this line of  argument and found  the upper limit to be $5\times 10^{10}$  M$_{\odot}$ for typical input parameters. Despite important role of the radiation pressure in the  SMBH discs, however, the analysis of \cite{king2016} is restricted to a gas-pressure dominated disc. \cite{king2016} argues that formation of a large radiation-dominated disc is not possible because of the thermal instability, however, as \cite{inay2016} stressed,  the implication of the thermal instability in this context is not yet understood. \cite{inay2016} investigated a similar problem using a disc model based on the \cite{Thompson2005} model, which includes star formation and the radiation pressure from the stars to maintain vertical hydrostatic  equilibrium against the gravity,  and, instead of the usual $\alpha$-prescription, the viscosity is determined by assuming that the radial Mach number is a constant input parameter. They demonstrated that growth of a SMBH is prevented by the accretion physics such that BH maximum mass is estimated to be between $10^{10}$ and $6\times 10^{10}$ solar masses.

Our analytical solutions enable us to estimate the SMBH maximum mass. We follow a similar approach to \cite{king2016}, in which the accretion process is suppressed if the self-gravitating radius shifts to a radius smaller than the innermost stable circular orbit, i.e. $r_{\rm ISCO}$. This radius is written as \citep{king2016},
\begin{equation}
r_{\rm ISCO} = f(a_{\rm s}) \frac{GM}{c^2},
\end{equation} 
where $f(a_{\rm s})$ is a function of the SMBH spin parameter $a_{\rm s}$, i.e. $f(a_{\rm s})^{1/2}[4-(3f(a_{\rm s})-2)^{1/2}]=3a_{\rm s}$ \citep[e.g.,][]{king2016}. For dimensionless Kerr parameter $a_{\rm s}=-1$, $0$, and $1$, we have $f(a_{\rm s})=9$, $6$, and $1$, respectively. Up to now we adopted a standard value  for the efficiency of the accretion ($\epsilon =0.1$), however, as we mentioned earlier, its value depends on the SMBH spin. \cite{bardeen}  showed that  
\begin{equation}\label{eq:effe}
\epsilon = 1- \sqrt{1-\frac{2}{3f(a_{\rm s})}}.
\end{equation}
Equation (\ref{eq:rsg}) gives self-gravitating radius as a function of the central mass and the accretion rate for $\epsilon=0.1$. We can now re-write this equation, but keeping the accretion efficiency $\epsilon$ as an input parameter given by Equation (\ref{eq:effe}). Thus,
\begin{equation}
\frac{r_{\rm sg}}{r_{\rm s}} \simeq 3 \times 10^3 \epsilon^{0.28} m^{-0.94} \dot{m}^{-0.28}.
\end{equation}
Our constraint to estimate SMBH maximum mass is $r_{\rm sg}\sim r_{\rm ISCO}$. Thus we have
\begin{equation}
M_{\rm max} = 1.05\times 10^{12} {\rm M}_{\odot} \epsilon^{0.28} \dot{m}^{-0.28} f^{-1.06}.
\end{equation}

\section{conclusion}
We considered both the radiation and the gas pressures in a gravitoturbulent accretion disc and pursued its consequences in detail. This approach permitted us to uniquely determine dimensionless stress parameter that is inversely proportional to the cooling rate. It prompted us to treat the onset of fragmentation of the disc in terms of a critical value for the stress parameter, rather than in terms of only Toomre parameter like most of the previous studies. The benefit of our model is that it has only two input parameters $M$ and $\dot{M}$, whereas previous related models incorporate three input parameter, i.e. $M$, $\dot{M}$ and $\alpha$.

Our study is different from previous works on properties of quasars discs in several ways:

1. In contrast to all previous works, we account for the dependence of dimensionless stress parameter on the properties of the disc  as a physical consequence of the thermal equilibrium of the disc.

2. While in previous works fragmentation of a quasar disc is explored via Toomre parameter, fragmentation of our gravitoturbulent disc with a Toomre parameter close to its threshold  is investigated once dimensionless stress parameter becomes larger than a threshold value.

3. In contrary to the previous works on  gravitational stability of the gravitoturbulent discs, in the presence of radiation pressure, this threshold of the instability is not fixed, and, it is a power-law function of the ratio of the gas pressure and the total pressure with an exponent about 1.7.

4. We also estimated the BH maximum mass using our analytical gravitoturbulent model and found that the maximum mass is consistent with the previous studies which rely on somewhat different models. 

Our results are also consistent with the numerical simulations. Morover, the present model enable us to investigate properties of gravitoturbulent quasar discs over a wider range of the input parameters. For instance,  \cite{goodman11} investigated gravitational stability of an accretion disc including dynamical and thermal roles of radiation for a central black hole with mass $10^8$ solar masses. We suggest this kind of  numerical simulations should be extended to the systems with a larger central masses even up to $10^9$  or $10^{10}$ solar masses. In these cases, our analytical solutions provide some physical insights about typical behaviors that one may expect. 

We note that a viscous formalism is not, in general, able to describe properties of a gravity-driven accretion disc due to inherently non-local nature of the gravitational force \citep[e.g.,][]{balbus99}. As long as the disc mass is much smaller than the central object, however, numerical simulations show that viscous approach is roughly adequate to model structure of a disc with the gravity-driven turbulence \citep{Lodato2004}. More precisely, once the disc-to-star mass ratio exceeds around 0.5 and/or the ratio of the disc thickness and the radial distance becomes larger than 0.1, a global treatment of the angular momentum transport is needed.  

Our estimate of boundary fragmentation relies on the simulations of self-gravitating discs which show that once the stress parameter reaches to its critical value $\alpha_{\rm C}$, then the disc is deemed to be subject to fragmentation. In the absence of radiation, there are considerable debates on the value of $\alpha_{\rm C}$. For specific heat ratios $\gamma =2$, \cite{gammie2001} showed that $\alpha_{\rm C}\simeq 0.07$ which is supported by numerical simulations of \cite{rice2005}, whereas \cite{rafikov2009} adopted $\alpha_{\rm C} \simeq 1$ in agreement with simulations of star formation processes ?. It has been argued that the disagreement may be due to numerical resolutions and recent high resolution simulations indicate that fragmentation occurs at smaller $\alpha_{\rm C}$. In the absence of radiation, we mainly adopted $\alpha_{\rm C}\simeq 1$   as our reference value and modified it as $\alpha_{\rm C} \simeq \beta^n$ when radiation and gas pressures are  considered. Some authors, however, showed that the fragmentation boundary in the absence of radiation occurs for much smaller value about 0.06 \citep[e.g.,][]{rice2005}. We think, therefore, if our modified critical stress parameter is introduced as $\alpha_{\rm C} \simeq 0.06 \beta^n $, then  self-gravitating radius shifts towards central BH for a given set of the input parameters and the effect of the slope $n$ appears to be same those cases which we explored  in Figures \ref{fig:f5} and \ref{fig:f6}.

It is also worth noting that boundary fragmentation in our model is not stochastic. Recent studies showed that turbulent discs are always gravitationally unstable in a probabilistic sense because of the chaotic nature of the turbulence \cite[e.g.,][]{hopkins13}. In the realm  of planet formation in PPDs, however, it was shown that the stochastic fragmentation does not modify self-gravitating radius by more than 20 percent \citep{young16}. On the other hand, formation of the first fragments in a PPD may cause further fragmentation at smaller radii than initially expected \citep[e.g.,][]{armi99,meru15}. Similar mechanisms may operate in quasar discs, though it has not been investigated yet.

\section*{Acknowledgements}
M. S. and F. K. thank the hospitality and support during their visit to Niels Bohr Institute \& Centre for Star and Planet Formation, Denmark, where part of this work has been done. M. S. is grateful to Iran Science Elites Federation for their support. 
S. D. is supported by a Marie-Curie Intra European Fellowship under the European Community's Seventh Framework Program FP7/2007-2013 grant agreement no 627008.




\bibliographystyle{mnras}
\bibliography{reference} 




\bsp	
\label{lastpage}
\end{document}